\documentclass[a4paper,11pt]{article}
\pdfoutput=1 

\usepackage{jheppub} 

\usepackage[T1]{fontenc} 

\title{\boldmath Stability in the higher derivative Abelian gauge field theory}

\author[]{J.L. Dai,\note{Corresponding author.}}

\affiliation[a]{Department of Physics, Zhejiang University, Hangzhou, 310027, P. R. China}
\affiliation[b]{Center of Mathematical Science, Zhejiang University, Hangzhou, 310027, P. R. China}

\emailAdd{daijlxy@126.com}

\abstract{We present the derivation of conserved tensors associated to higher-order symmetries in the higher derivative Maxwell Abelian gauge field theories. In our model, the wave operator of the higher derived theory is a $n$-th order polynomial expressed in terms of the usual Maxwell operator.  Any symmetry of the primary wave operator gives rise to a collection of independent higher-order symmetries of the field equations which thus leads to a series of independent conserved quantities of derived system. In particular, by the extension of Noether's theorem, the spacetime translation invariance of the Maxwell primary operator results in the series of conserved second-rank tensors which includes the standard canonical energy-momentum tensors. Although this canonical energy is unbounded from below, by introducing a set of parameters, the other conserved tensors in the series can be bounded which ensure the stability of the higher derivative dynamics. In addition, with the aid of auxiliary fields, we successfully obtain the relations between the roots decomposition of characteristic polynomial of the wave operator and the conserved energy-momentum tensors within the context of another equivalent lower-order representation. Under the certain conditions, the 00-component of the linear combination of these conserved quantities is bounded and by this reason, the original derived theory is considered stable. Finally, as an instructive example, we discuss the third-order derived system and analyze extensively the stabilities in different cases of roots decomposition.}

\begin{document}
\maketitle
\flushbottom

\section{Introduction}

The research of higher-order derivative systems dates back to the nineteenth century in Ostrogradsky's pioneering work~\cite{1} and is still actively studied nowadays such as in the areas of effective low energy theories, astrophysical and cosmological behaviors and the modified gravities~\cite{2,3,4,5,6,7,8,9}. This promising approach was first introduced in field theories to get rid of the infinities associated to point particles~\cite{10,11,12} and later, Pais and Uhlenbeck proposed a class of classical higher derivative harmonic oscillators~\cite{13}. In these examples, the Lagrangian containing higher derivative terms was been very attractive due to its nice ultraviolet behaviour and will result in a renormalizable quantum field theory~\cite{14}. However, such Lagrangians yield higher-order equations of motion, which require more initial
conditions than in usual dynamical systems and a standard framework for dealing with these theories on Hamiltonian level is provided by Ostrogradski canonical approach~\cite{15,16,17,18}. Unfortunately, the Hamiltonian functions obtained in such a way contain terms linear in momenta and are almost always unbounded from below. Therefore with the presence of higher derivatives, the systems turn out to be unstable~\cite{19,20,21} and moreover, the existence of unbounded kinetic terms inevitably lead to runaway solutions if interactions are turned on.

In order to circumvent these problems in a physical allowed sector, various motivations and techniques have been put forward to avoid the Ostrogradsky ghosts in different higher derivative models~\cite{22,23,24,25,26,27,28,29,30}. For instance, at least in the Pais-Uhlenbeck's harmonic oscillator which is served as a toy model to understand several
important issues related to Ostrogradsky instabilities, Raidal and Veermae advised that for the purpose of the energy spectrum of the theory be bounded,  the ghost degrees of freedom should be necessarily complex~\cite{31}. In this sense, the resulting complex system can be consistently quantized using the rules of canonical quantisation which possesses all good properties of the known quantum physics including the positive definite Hamiltonian. In the usual Lee-Wick theories, from the viewpoint of polynomials with complex conjugate poles, it is possible to construct a unitary $S$-matrix of gravitational excitations to remove the negative effects of the ghosts~\cite{32,33}. On the other hand, in contrast to the classical systems, quantizing the higher derivative dynamics imposes even more constraints. It is thought that higher-order theories would possess propagators having poles with non-positive residues which lead to the appearance of ghost states. At the quantum level, these ghost states have non-positive norms and due to this, they will violate the causality and spoil the unitarity evolution of the quantum theory which is unacceptable physically. By introducing form-factors with an analytic dependence on the propagating momenta ~\cite{34}, we are able to avoid the unphysical ghosts and this method will preserve all fundamental properties of a quantum field theory. Furthermore, in the non-Hermitian, $\mathcal{PT}$-symmetric model, i.e., symmetric under combined parity reflection and time reversal, it is essential to modify the dynamical inner product instead of using the standard Dirac inner product~\cite{35,36,37,38}. In this manner, we explicitly obtain the self-adjoint Hamiltonian and its ghost state is reinterpreted as an ordinary quantum state with positive $\mathcal{PT}$ norm which gives rise to the standard probabilistic interpretation.

In the~\cite{39}, a special class of linear higher derivative systems is discussed as an alternative approach to the problem of Ostrogradsky instability. To be more precise, the operators of the dynamic equations in these theories are supposed to be factorable in terms of a pair of different second-order operators satisfying some certain conditions. In this way, with the help of auxiliary fields, it is possible to establish two equivalent systems which may be thought of as two different representations of the same theory. Then the Noether's theorem tells us that if the action functional is preserved under the spacetime translations, the system is equipped with canonical energy-momentum tensors and the 00-component is of particular importance since it has the sense of energy density and will lead to the energy conservation law. Especially, for the models of factorable type, applying the Noether's theorem, we are capable of acquiring two family of integrals of motion which may be either bounded or unbounded depending on the specific values of parameters. As is explained in~\cite{39}, the stability of the higher derivative system can be ensured if the 00-component in this family is positive definite even if the Noether's canonical energy is unbounded. So far, the efforts of this approach have been focused mostly on various known factorable models such as Pais-Uhlenbeck's harmonic oscillators, higher derivative scalar fields and Podolsky's generalized electrodynamics.

After that, a more general and systematic method was carried out as a guide to investigate the stabilities in a wide class of higher derivative systems named derived type theories~\cite{40,41,42,43,44}. Generally speaking, these derived theories are based on simpler free primary models whose equations of motion only involve first and second order differential operators without higher derivatives. In this setting, the wave operator which determines the dynamic equations of the higher derivative systems is a polynomial in terms of the primary wave operator in the lower-order free theory. Then, every symmetry of primary theory enables us to construct $n$-parametric series of symmetries of the derived theory if the order of the characteristic polynomial of the wave operator is $n$. More importantly, these symmetries are connected to $n$ independent conserved quantities from the perspective of more general correspondence between symmetries and conservation laws which is established by the Noether's theorem as well as the Lagrange anchor~\cite{45,46}. Especially, when the primary wave operator commutes with the spacetime translation generators, the derived theories have $n$-parametric series of conserved second-rank tensors $(T_{k})^{\nu}_{\mu},k=0,1,...,n-1$ and in particular, the $k=0$ term corresponds to the usual canonical energy $(T_{0})^{0}_{0}$ of the higher derivative systems. Now although the canonical energy is unbounded due to the nature of higher derivatives, the linear combination of these tensors $(T_{k})^{\nu}_{\mu}$ may give rise to bounded conserved charge which will stabilize the classical dynamics of derived model at free level, which also persists at quantum level. Moreover, as demonstrated at length in ~\cite{40,41}, when these conserved tensors are bounded in the free theory, the inclusion of consistent interactions will not spoil the stability of the coupling systems, at least at perturbative level.

The paper is organized as follows. In section 2, we start by describing the Lagrangians for the higher derivative Maxwell gauge field theories by means of general wave operators. Subsequently, we give a detailed
derivation of series of second-rank conserved tensors from the higher-order symmetries and investigate the issue of the stability in this derived system. In section 3, according to the different root decompositions of the characteristic polynomials, we set up the formulae of the conserved tensors associated with real and complex roots with the aid of auxiliary fields. Then as an application, section 4 is devoted to the full analysis of the stabilities in the third-order derived system. The final section of this paper includes some concluding remarks and discussions.

\section{Higher derivative Abelian gauge theory}
\subsection{Conserved tensors}
Let us start with the Lagrangian density of usual Maxwell electromagnetic theory which is described by the gauge fields $A_{\mu}$ in (1+3)-dimensional spacetime as follows
\begin{equation}
\begin{aligned}
S=-\frac{1}{4}\int F^{\mu\nu}F_{\mu\nu}d^{4}x
\end{aligned}
\end{equation}
here the metric is $g_{\mu\nu}=\mathrm{diag}(1,-1,-1,-1)$ which can be used to rise and lower the multi-indices. The dynamic equations of motion of (2.1) are simply present as
\begin{equation}
\begin{aligned}
\partial_{\mu}F^{\mu\nu}=0
\end{aligned}
\end{equation}
if we set
\begin{equation}
\begin{aligned}
W_{\mu\nu}=\delta_{\mu\nu}\square-\partial_{\nu}\partial_{\mu}
\end{aligned}
\end{equation}
as the primary wave operator, then (2.3) defines the primary free field equation~\cite{41}
\begin{equation}
\begin{aligned}
W_{\mu\nu}A^{\nu}=0
 \end{aligned}
\end{equation}
based on this primary model, the most general Lagrangian density of the higher-order extensions of Maxwell gauge theory is given by
\begin{equation}
\begin{aligned}
S=\int d^{4}xA_{\mu}M_{\mu\nu}A^{\nu}
\end{aligned}
\end{equation}
here $M$ is termed as wave operator which is a polynomial in the formal variable $W$
\begin{equation}
\begin{aligned}
M=a_{n}W^{n}+......+a_{2}W^{2}+a_{1}W+a_{0}
\end{aligned}
\end{equation}
and the equation of motion contains terms up to the 2$n$-th time derivative
\begin{equation}
\begin{aligned}
\sum_{l=0}^{n}a_{l}W^{l}_{\mu\nu}A^{\nu}=0
\end{aligned}
\end{equation}

 It is well known that the symmetry of a field theory plays a very significant role in modern physics and it has been regarded as one of the most powerful tools to analyze the behaviors of the physical gauge systems. In particular, the simplest possible and useful symmetry of the free field theory is the translation invariance which means that the translation generators $\partial_{\mu}$ commute with the primary wave operator in the form of
\begin{equation}
\begin{aligned}
\left[\partial_{\mu},W\right]=0
\end{aligned}
\end{equation}
this assumption implies that the derived theory (2.5) enjoys the following higher-order symmetries
\begin{equation}
\begin{aligned}
\delta_{\varepsilon}A^{\mu}=\varepsilon^{\nu}\partial_{\nu}(W^{k}A)^{\mu},\quad \quad \quad k=0,1,...,n-1
\end{aligned}
\end{equation}
 especially, $k=0$ corresponds to the spacetime translations invariance of the action of the derived model. Now the Noether's theorem tells us that each continuous symmetry of the action (2.9) determines a conserved quantity~\cite{40,41}
\begin{equation}
\begin{aligned}
\partial_{\mu}(\Theta^{k})^{\mu}_{\nu}=(\partial_{\nu}(W^{k}A)^{\mu})(MA)_{\mu}
\end{aligned}
\end{equation}
in the above expression, we obtain $n$ independent conserved tensors and the $k=0$ term corresponds to the Noether's canonical energy-momentum tensors, while the other $k\geq1$ terms are different conserved tensors connected to the higher-order symmetries of the gauge fields $A_{\mu}$.

To find out the explicit expressions of $(\Theta^{k})^{\mu}_{\nu}$, at first, a simple calculation shows that
\begin{equation}
\begin{aligned}
(W^{k}A)_{\mu}=\square^{k-1}\partial^{\rho}F_{\rho\mu},\quad\quad k\geq1
\end{aligned}
\end{equation}
for convenience, we define
\begin{equation}
\begin{aligned}
\square^{-1}\partial^{\rho}F_{\rho\mu}:=A_{\mu}
\end{aligned}
\end{equation}
thus in this way, the dynamical equation of motion (2.7) turns out to be
\begin{equation}
\begin{aligned}
\sum_{l=0}^{n}a_{l}\square^{l-1}\partial_{\mu}F^{\mu\nu}=0
\end{aligned}
\end{equation}
subsequently, if $l=k$ in (2.10), it is easy to see that
\begin{equation}
\begin{aligned}
(\partial_{\nu}(W^{k}A)^{\mu})(W^{k}A)_{\mu}=\frac{1}{2}\partial_{\nu}(\square^{k-1}\partial^{\rho}F_{\rho\lambda}\square^{k-1}\partial_{\tau}F^{\tau\lambda})
\end{aligned}
\end{equation}
on the other hand, when $l\geq k+1$, in view of
\begin{equation}
\begin{aligned}
&\partial_{\nu}\partial^{\rho}F_{\rho\mu}-\partial_{\mu}\partial^{\rho}F_{\rho\nu}=\square F_{\nu\mu}
\end{aligned}
\end{equation}
there is no difficulty in evaluating
\begin{equation}
\begin{aligned}
&(\partial_{\nu}(W^{k}A)^{\mu})(W^{l}A)_{\mu}\\
=&(\partial_{\nu}\square^{k-1}\partial^{\rho}F_{\rho\mu})\square^{l-1}\partial_{\lambda}F^{\lambda\mu}\\
=&\square^{k}F_{\nu\mu}\square^{l-1}\partial_{\lambda}F^{\lambda\mu}+(\square^{k-1}\partial_{\mu}\partial^{\rho}F_{\rho\nu})\square^{l-1}\partial_{\lambda}F^{\lambda\mu}\\
=&\partial_{\lambda}(\square^{k}F_{\nu\mu}\square^{l-1} F^{\lambda\mu})-(\square^{k}\partial_{\lambda}F_{\nu\mu})\square^{l-1} F^{\lambda\mu}+\partial_{\mu}(\square^{k-1}\partial^{\rho}F_{\rho\nu}\square^{l-1}\partial_{\lambda}F^{\lambda\mu})\\
\end{aligned}
\end{equation}
then making using of
\begin{equation}
\begin{aligned}
\partial_{\lambda}F_{\nu\mu}+\partial_{\nu}F_{\mu\lambda}+\partial_{\mu}F_{\lambda\nu}=0
\end{aligned}
\end{equation}
and taking into account of the symmetry among the indices $\lambda,\mu$, we infer that
\begin{equation}
\begin{aligned}
-(\square^{k}\partial_{\lambda}F_{\nu\mu})\square^{l-1} F^{\lambda\mu}=\frac{1}{2}(\square^{k}\partial_{\nu}F_{\mu\lambda})\square^{l-1} F^{\lambda\mu}\\
\end{aligned}
\end{equation}
at this stage, if $l=k+1$ we are thus led to
\begin{equation}
\begin{aligned}
\frac{1}{2}(\square^{k}\partial_{\nu}F_{\mu\lambda})\square^{k} F^{\lambda\mu}=\frac{1}{4}\partial_{\nu}(\square^{k}F_{\mu\lambda}\square^{k} F^{\lambda\mu})
\end{aligned}
\end{equation}
while $l\geq k+2$, applying the general procedure of integration by parts together with (2.17), we simply have
\begin{equation}
\begin{aligned}
&\frac{1}{2}(\square^{k}\partial_{\nu}F_{\mu\lambda})\square^{l-1} F^{\lambda\mu}\\
=&\frac{1}{2}\partial_{\nu}(\square^{k}F_{\mu\lambda}\square^{l-1} F^{\lambda\mu})-\frac{1}{2}\square^{k}F_{\mu\lambda}\square ^{l-1} \partial_{\nu}F^{\lambda\mu}\\
=&\frac{1}{2}\partial_{\nu}(\square^{k}F_{\mu\lambda}\square^{l-1} F^{\lambda\mu})+\square^{k}F^{\mu\lambda}\square^{l-1} \partial_{\mu}F_{\nu\lambda}\\
=&\frac{1}{2}\partial_{\nu}(\square^{k}F_{\mu\lambda}\square^{l-1} F^{\lambda\mu})+\partial_{\mu}(\square^{k}F^{\mu\lambda}\square^{l-1} F_{\nu\lambda})-(\partial_{\mu} \square^{k}F^{\mu\lambda})\square^{l-1} F_{\nu\lambda}\\
\end{aligned}
\end{equation}
as well as
\begin{equation}
\begin{aligned}
&-(\square^{k}\partial_{\mu} F^{\mu\lambda})\square^{l-1} F_{\nu\lambda}\\
=&-(\square^{k}\partial_{\mu} F^{\mu\lambda})\square^{l-2}(\partial_{\nu}\partial^{\rho}F_{\rho\lambda}-\partial_{\lambda}\partial^{\rho}F_{\rho\nu})\\
=&-\partial_{\nu}(\square^{k}\partial_{\mu} F^{\mu\lambda}\square^{l-2}\partial^{\rho}F_{\rho\lambda})+(\square^{k}\partial_{\nu}\partial_{\mu} F^{\mu\lambda})\square^{l-2}\partial^{\rho}F_{\rho\lambda}+\partial_{\lambda}(\square^{k}\partial_{\mu} F^{\mu\lambda}\square^{l-2}\partial^{\rho}F_{\rho\nu})
\end{aligned}
\end{equation}
furthermore, using (2.15), after a straightforward computation we get
\begin{equation}
\begin{aligned}
&(\square^{k}\partial_{\nu}\partial_{\mu} F^{\mu\lambda})\square^{l-2}\partial^{\rho}F_{\rho\lambda}\\
=&(\square^{k+1}F_{\nu\lambda})\square^{l-2}\partial_{\rho}F^{\rho\lambda}+(\square^{k}\partial_{\lambda}\partial^{\mu} F_{\mu\nu})\square^{l-2}\partial_{\rho}F^{\rho\lambda}\\
=&\partial_{\rho}(\square^{k+1}F_{\nu\lambda}\square^{l-2}F^{\rho\lambda})-(\square^{k+1}\partial_{\rho } F_{\nu\lambda})\square^{l-2}F^{\rho\lambda}+\partial_{\lambda}(\square^{k}\partial^{\mu} F_{\mu\nu}\square^{l-2}\partial_{\rho}F^{\rho\lambda})\\
=&\partial_{\rho}(\square^{k+1}F_{\nu\lambda}\square^{l-2}F^{\rho\lambda})+\frac{1}{2}\square^{k+1}\partial_{\nu}F_{\lambda\rho}\square^{l-2}F^{\rho\lambda}+\partial_{\lambda}(\square^{k}\partial^{\mu} F_{\mu\nu}\square^{l-2}\partial_{\rho}F^{\rho\lambda})
\end{aligned}
\end{equation}
comparing (2.20) to (2.22) and employing a recursive algorithm, we acquire the following equation
\begin{equation}
\begin{aligned}
&\frac{1}{2}(\square^{k}\partial_{\nu}F_{\mu\lambda})\square^{l-1} F^{\lambda\mu}\\
=&\sum_{i=0}^{l-k-2}(\frac{1}{2}\partial_{\nu}(\square^{k+i}F_{\mu\lambda}\square^{l-1-i} F^{\lambda\mu})+2\partial_{\mu}(\square^{k+i}F^{\mu\lambda}\square^{l-1-i} F_{\nu\lambda})-\partial_{\nu}(\square^{k+i}\partial_{\mu} F^{\mu\lambda}\square^{l-2-i}\partial^{\rho}F_{\rho\lambda})\\
&+2\partial_{\mu}(\square^{k+i}\partial_{\lambda} F^{\lambda\mu}\square^{l-2-i}\partial^{\rho}F_{\rho\nu}))+\frac{1}{2}\square^{l-1}\partial_{\nu}F^{\lambda\rho}\square^{k}F_{\rho\lambda}
\end{aligned}
\end{equation}
particularly, in the above derivation of (2.23), we have used the identities
\begin{equation}
\begin{aligned}
&\sum_{i=0}^{l-k-2}\partial_{\rho}(\square^{k+1+i}F_{\nu\lambda}\square^{l-2-i}F^{\rho\lambda})=\sum_{i=0}^{l-k-2}\partial_{\rho}(\square^{k+i}F^{\rho\lambda}\square^{l-1-i}F_{\nu\lambda}),\\
&\sum_{i=0}^{l-k-2}\partial_{\mu}(\square^{k+i}\partial^{\lambda}F_{\lambda\nu}\square^{l-2-i}\partial_{\rho}F^{\rho\mu})=\sum_{i=0}^{l-k-2}\partial_{\mu}(\square^{k+i}\partial_{\lambda} F^{\lambda\mu}\square^{l-2-i}\partial^{\rho}F_{\rho\nu})
\end{aligned}
\end{equation}

According to these results, we are able to formulate the higher-order conserved tensors in (2.10) in the form of
\begin{equation}
\begin{aligned}
(\Theta^{k})^{\mu}_{\nu}=\sum_{l=k+1}^{n}a_{l}(t_{1}^{k,l})^{\mu}_{\nu}+\frac{1}{2}\delta^{\mu}_{\nu}a_{k}(\square^{k-1}\partial^{\rho}F_{\rho\lambda}\square^{k-1}\partial_{\tau}F^{\tau\lambda})+\sum_{l=0}^{k-1}a_{l}(t_{2}^{k,l})^{\mu}_{\nu}
\end{aligned}
\end{equation}
here
\begin{equation}
\begin{aligned}
l=k+1:(t_{1}^{k,l})^{\mu}_{\nu}=&\square^{k}F_{\nu\lambda}\square^{k} F^{\mu\lambda}-\frac{1}{4}\delta^{\mu}_{\nu}(\square^{k}F_{\rho\lambda})\square^{k} F^{\rho\lambda}+\square^{k-1}\partial^{\rho}F_{\rho\nu}\square^{k}\partial_{\tau}F^{\tau\mu},\\
l\geq k+2:(t_{1}^{k,l})^{\mu}_{\nu}=&\square^{k}F_{\nu\lambda}\square^{l-1} F^{\mu\lambda}-\frac{1}{4}\delta^{\mu}_{\nu}(\square^{k}F_{\rho\lambda})\square^{l-1} F^{\rho\lambda}+\sum_{i=0}^{l-k-2}(\square^{k+i}F^{\mu\lambda}\square^{l-1-i} F_{\nu\lambda}\\
&-\frac{1}{4}\delta^{\mu}_{\nu}\square^{k+i}F_{\rho\lambda}\square^{l-1-i} F^{\rho\lambda}+\square^{k+i}\partial_{\lambda} F^{\lambda\mu}\square^{l-2-i}\partial^{\rho}F_{\rho\nu}\\
&-\frac{1}{2}\delta^{\mu}_{\nu}\square^{k+i}\partial_{\tau} F^{\tau\lambda}\square^{l-2-i}\partial^{\rho}F_{\rho\lambda})+\square^{k-1}\partial^{\rho}F_{\rho\nu}\square^{l-1}\partial_{\tau}F^{\tau\mu}\\
\end{aligned}
\end{equation}
analogously, when $l\leq k-1$, after integration by parts
\begin{equation}
\begin{aligned}
(\partial_{\nu}(W^{k}A)^{\mu})(W^{l}A)_{\mu}=\partial_{\nu}((W^{k}A)^{\mu}(W^{l}A)_{\mu})-(W^{k}A)^{\mu}\partial_{\nu}(W^{l}A)_{\mu}
\end{aligned}
\end{equation}
and taking a similar tactic in the case of $l>k$, it is not difficult to re-express the $(W^{k}A)^{\mu}\partial_{\nu}(W^{l}A)_{\mu}$ as total derivative terms which give rise to the exact expressions of $(t_{2}^{k,l})^{\mu}_{\nu}$, these are
\begin{equation}
\begin{aligned}
l=k-1:(t_{2}^{k,l})^{\mu}_{\nu}=&\delta^{\mu}_{\nu}\square^{k-1}\partial^{\rho}F_{\rho\lambda}\square^{k-2}\partial_{\tau}F^{\tau\lambda}-\square^{k-1}F_{\nu\lambda}\square^{k-1} F^{\mu\lambda}+\frac{1}{4}\delta^{\mu}_{\nu}(\square^{k-1}F_{\rho\lambda})\square^{k-1}F^{\rho\lambda}\\
&-\square^{k-2}\partial^{\rho}F_{\rho\nu}\square^{k-1}\partial_{\tau}F^{\tau\mu},\\
l\leq k-2:(t_{2}^{k,l})^{\mu}_{\nu}=&\delta^{\mu}_{\nu}\square^{k-1}\partial^{\rho}F_{\rho\lambda}\square^{l-1}\partial_{\tau}F^{\tau\lambda}-\square^{l}F_{\nu\lambda}\square^{k-1} F^{\mu\lambda}+\frac{1}{4}\delta^{\mu}_{\nu}(\square^{l}F_{\rho\lambda})\square^{k-1} F^{\rho\lambda}\\
&-\sum_{i=0}^{k-l-2}(\square^{l+i}F^{\mu\lambda}\square^{k-1-i} F_{\nu\lambda}-\frac{1}{4}\delta^{\mu}_{\nu}\square^{l+i}F_{\rho\lambda}\square^{k-1-i} F^{\rho\lambda}+\square^{l+i}\partial_{\lambda} F^{\lambda\mu}\square^{k-2-i}\partial^{\rho}F_{\rho\nu}\\
&-\frac{1}{2}\delta^{\mu}_{\nu}\square^{l+i}\partial_{\tau} F^{\tau\lambda}\square^{k-2-i}\partial^{\rho}F_{\rho\lambda})-\square^{l-1}\partial^{\rho}F_{\rho\nu}\square^{k-1}\partial_{\tau}F^{\tau\mu}\\
\end{aligned}
\end{equation}

\subsection{Stability}
Once obtained the explicit expressions of $(\Theta^{k})^{\mu}_{\nu}$, we wish to investigate the problem of stability in (2.5) by introducing $n$ independent parameters
\begin{equation}
\begin{aligned}
\beta_{0},\quad \beta_{1},\quad......,\quad  \beta_{n-1}
\end{aligned}
\end{equation}
and the total series of second-rank energy-momentum tensors of the derived theory under current study reads as~\cite{40,41}
\begin{equation}
\begin{aligned}
\Theta^{\mu}_{\nu}(A,\beta)=\sum_{k=0}^{n-1}\beta_{k}(\Theta^{k})^{\mu}_{\nu}
\end{aligned}
\end{equation}
this family of conserved tensors includes the canonical energy-momentum $(\Theta^{0})^{\mu}_{\nu}$ of the derived model (2.5) when $\beta_{0}=1, \beta_{1}=......=\beta_{n-1}=0$, though it is always unbounded and the other conserved quantities originate from the higher-order symmetries in the set (2.9). In the light of this, the 00-component of this conserved tensor has the meanings of the energy density of the higher derivative system and the total energy of the derived theory is provided by the integral
\begin{equation}
\begin{aligned}
E=\int d^{4}x\Theta^{0}_{0}
\end{aligned}
\end{equation}
as far as the issue of stability of the higher derivative model is concerned,  our strategy is to guarantee the positive definite of the total energy which can be achieved by the requirement $\Theta^{0}_{0}\geq0$.

In the present situation, choosing $\mu=\nu=0$ in (2.25) and with the aid of metric $g_{\mu\nu}=\mathrm{diag}(1,-1,-1,-1)$, it is possible to cast the 00-component of $(\Theta^{k})^{\mu}_{\nu}$ in the form
\begin{equation}
\begin{aligned}
(\Theta^{k})^{0}_{0}=&-\frac{1}{4}\sum_{l=k+1}^{n}a_{l}\square^{k}F_{\rho\lambda}\square^{l-1} F_{\rho\lambda}+\frac{1}{4}\sum_{l=0}^{k-1}a_{l}\square^{l}F_{\rho\lambda}\square^{k-1} F_{\rho\lambda}\\
&-\frac{1}{2}\sum_{l=k+2}^{n}\sum_{i=0}^{l-k-2}a_{l}(\frac{1}{2}\square^{k+i}F_{\rho\lambda}\square^{l-1-i}F_{\rho\lambda}-\square^{k+i}\partial^{\mu} F_{\mu\lambda}\square^{l-2-i}\partial^{\rho}F_{\rho\lambda})\\
&+\frac{1}{2}\sum_{l=0}^{k-2}\sum_{i=0}^{k-l-2}a_{l}(\frac{1}{2}\square^{l+i}F_{\rho\lambda}\square^{k-1-i}F_{\rho\lambda}-\square^{l+i}\partial^{\mu} F_{\mu\lambda}\square^{k-2-i}\partial^{\rho}F_{\rho\lambda})\\
&+\sum_{l=k+1}^{n}a_{l}\square^{k-1}\partial^{\rho}F_{\rho0}\square^{l-1}\partial_{\tau}F^{\tau0}+\sum_{l=0}^{k-1}a_{l}\square^{k-1}\partial^{\rho}F_{\rho\lambda}\square^{l-1}\partial_{\tau}F^{\tau\lambda}\\
&-\sum_{l=0}^{k-1}a_{l}\square^{l-1}\partial^{\rho}F_{\rho0}\square^{k-1}\partial_{\tau}F^{\tau0}+\frac{1}{2}a_{k}(\square^{k-1}\partial^{\rho}F_{\rho\lambda}\square^{k-1}\partial_{\tau}F^{\tau\lambda})
\end{aligned}
\end{equation}
notice that using the equations of motion (2.13) and setting $\nu=0$, we have
\begin{equation}
\begin{aligned}
\sum_{l=k+1}^{n}a_{l}\square^{l-1}\partial_{\tau}F^{\tau0}=-a_{k}\square^{k-1}\partial_{\tau}F^{\tau0}-\sum_{l=0}^{k-1}a_{l}\square^{l-1}\partial_{\tau}F^{\tau0}
\end{aligned}
\end{equation}
inserting this relation back into $(\Theta^{k})^{0}_{0}$, one can check that the last four terms in (2.32) could be rewritten as follows
\begin{equation}
\begin{aligned}
-\sum_{l=0}^{k-1}a_{l}\square^{k-1}\partial^{\rho}F_{\rho\lambda}\square^{l-1}\partial^{\tau}F_{\tau\lambda}-\frac{1}{2}a_{k}(\square^{k-1}\partial^{\rho}F_{\rho\lambda}\square^{k-1}\partial^{\tau}F_{\tau\lambda})
\end{aligned}
\end{equation}
which permits us to express the total energy density in a more concise and compact way
\begin{equation}
\begin{aligned}
\Theta^{0}_{0}=\sum_{k=0}^{n-1}\beta_{k}(\Theta^{k})^{0}_{0}=\sum_{i,j=0}^{n-1}(A_{ij}(a,\beta)\square^{i}F_{\rho\lambda}\square^{j} F_{\rho\lambda}+B_{ij}(a,\beta)\square^{i}\partial^{\rho}F_{\rho\lambda}\square^{j}\partial^{\tau}F_{\tau\lambda})
\end{aligned}
\end{equation}
here $A_{ij}(a,\beta),B_{ij}(a,\beta)$ are polynomial functions of the variables $a_{l},\beta_{k}$ which can be determined from (2.30),(2.32) and (2.34). Inspecting the above formulas, we observe that the 00-component of the energy density is a quadratic form of the formal variables $\square^{i}F_{\rho\lambda},\square^{i}\partial^{\rho}F_{\rho\lambda}$, therefore $\Theta^{0}_{0}$ is positive if
\begin{equation}
\begin{aligned}
A_{ij}(a,\beta),\quad  B_{ij}(a,\beta) \quad  \mathrm{are \quad  all \quad positive \quad definite\quad  matrices}
\end{aligned}
\end{equation}
in other words, once the coefficients $a_{l}$ and parameters $\beta_{k}$ satisfy these positive definite conditions, the original free Abelian derived theory (2.5) admits bounded conserved quantities which thus is considered stable, though its canonical energy is unbounded from below.

\section{Root decompositions}
It is well known that every  polynomial in principle can be formulated in terms of its roots and in this section, we want to establish the relations between the conserved tensors and the structure of roots of characteristic polynomial of the higher derivative Maxwell derived theory. In fact, under certain assumptions about the roots, it is possible to obtain the bounded 00-component of the conserved quantities which may not been seen directly from the general expression (2.6). In order to do so, we suppose the wave operator of the derived theory has the following decomposition structure
\begin{equation}
\begin{aligned}
M=\sum_{l=0}^{n}a_{l}W^{l}=\prod_{i=1}^{p}(W-\lambda_{i})^{p_{i}}\prod_{j=1}^{q}(W^{2}-(\omega_{j}+\bar{\omega}_{j})W+\omega_{j}\bar{\omega}_{j})^{q_{j}}
\end{aligned}
\end{equation}
without loss of generality, we assume $a_{n}=1$ and the numbers $\lambda_{i},\omega_{j},\bar{\omega}_{j}$ label different real roots and complex roots. In addition, the numbers $p_{i},q_{j}$ are the corresponding multiplicities and the indices $p,q$ satisfy the condition
\begin{equation}
\begin{aligned}
\sum_{i=1}^{p}p_{i}+2\sum_{j=1}^{q}q_{j}=n
\end{aligned}
\end{equation}

\subsection{Real roots case}
Based on the above decomposition, firstly for every real root $\lambda_{i}$ and complex conjugate roots $\omega_{j},\bar{\omega}_{j}$, it is useful to define the new dynamic fields to absorb the higher derivatives of the original fields
\begin{equation}
\begin{aligned}
\xi_{k}&=\prod_{i=1,\atop i\neq k}^{p}(W-\lambda_{i})^{p_{i}}\prod_{j=1}^{q}(W^{2}-(\omega_{j}+\bar{\omega}_{j})W+\omega_{j}\bar{\omega}_{j})^{q_{j}}A,\\ \eta_{k}&=\prod_{i=1}^{p}(W-\lambda_{i})^{p_{i}}\prod_{j=1,\atop j\neq k}^{q}(W^{2}-(\omega_{j}+\bar{\omega}_{j})W+\omega_{j}\bar{\omega}_{j})^{q_{j}}A
\end{aligned}
\end{equation}
 by this construction, when the original fields $A_{\mu}$ are subject to the higher derivative field equations (2.7), these new component fields of course
fulfill the lower-order derived equations
\begin{equation}
\begin{aligned}
(W-\lambda_{i})^{p_{i}}\xi_{i}=0,\quad \quad (W^{2}-(\omega_{j}+\bar{\omega}_{j})W+\omega_{j}\bar{\omega}_{j})^{q_{j}}\eta_{j}=0
\end{aligned}
\end{equation}
for $i=1,2,...,p$ and $j=1,2,...,q$. Moreover, we observe that these dynamic equations also come from the following action functional
\begin{equation}
\begin{aligned}
S=\int d^{4}x\left[\sum_{i=1}^{p}\xi_{i}(W-\lambda_{i})^{p_{i}}\xi_{i}+\sum_{j=1}^{q}\eta_{j}(W^{2}-(\omega_{j}+\bar{\omega}_{j})W+\omega_{j}\bar{\omega}_{j})^{q_{j}}\eta_{j}\right]
\end{aligned}
\end{equation}
at this point, it is an easy exercise to show that the relations (3.3) allow us to establish one-to-one correspondence between the solutions to the higher derivative Maxwell gauge theory (2.5) and the lower-order dynamical system (3.5). In other words, these two systems are equivalent and can be viewed as two different representations of the same theory which are usually called $A$- and $\xi_{i}\eta_{i}$-representations.

On the other hand, noting that all the fields $\xi_{i}, \eta_{j}$ are independent degrees of freedom which means the action
functional $S$ possesses the following variational symmetries
\begin{equation}
\begin{aligned}
\delta_{\varepsilon}\xi_{i}=\varepsilon^{\mu}\partial_{\mu}(W-\lambda_{i})^{k}\xi_{i},\quad \quad \quad k=0,1,...,p_{i}-1
\end{aligned}
\end{equation}
for $\xi_{i}$ and under this condition, it immediately follows that the spacetime translation invariance of the primary operator $W$ gives us a series of conserved tensors
\begin{equation}
\begin{aligned}
\partial_{\mu}(T^{k}_{i})^{\mu}_{\nu}=\partial_{\nu}((W-\lambda_{i})^{k}\xi_{i})(W-\lambda_{i})^{p_{i}}\xi_{i},\quad \quad i=1,2,...,p
\end{aligned}
\end{equation}
then a direct calculation leads to
\begin{equation}
\begin{aligned}
&\partial_{\nu}((W-\lambda_{i})^{k}\xi_{i})(W-\lambda_{i})^{p_{i}}\xi_{i}=\sum_{j=0} ^{k}\sum_{l=0} ^{p_{i}} C_{k}^{j}(-\lambda_{i})^{k-j}\partial_{\nu}W^{j}\xi_{i}C_{p_{i}}^{l}(-\lambda_{i})^{p_{i}-l}W^{l}\xi_{i}\\
\end{aligned}
\end{equation}
making using of (2.25), we are able to formulate the second-rank conserved tensors $(T^{k}_{i})^{\mu}_{\nu}$ in the form of
\begin{equation}
\begin{aligned}
(T^{k}_{i})^{\mu}_{\nu}(\xi_{i})=&\sum_{l=0} ^{p_{i}}(\sum_{j=0}^{l-1}C_{k}^{j}C_{p_{i}}^{l}(-\lambda_{i})^{k+p_{i}-j-l}(\tilde{t}_{1}^{j,l})^{\mu}_{\nu}+\sum_{j=l+1}^{k}C_{k}^{j}C_{p_{i}}^{l}(-\lambda_{i})^{k+p_{i}-j-l}(\tilde{t}_{2}^{j,l})^{\mu}_{\nu})\\
&+\frac{1}{2}\sum_{l=0} ^{p_{i}}C_{k}^{l}C_{p_{i}}^{l}(-\lambda_{i})^{k+p_{i}-2l}\delta^{\mu}_{\nu}(\square^{l-1}\partial^{\rho}\tilde{F}^{i}_{\rho\lambda}\square^{l-1}\partial_{\tau}\tilde{F}_{i}^{\tau\lambda})\\
\end{aligned}
\end{equation}
for convenience, here we adopt the notations
\begin{equation}
\begin{aligned}
(\tilde{t}_{1}^{j,l})^{\mu}_{\nu}=(t_{1}^{j,l})^{\mu}_{\nu}(\xi_{i}),\quad (\tilde{t}_{2}^{j,l})^{\mu}_{\nu}=(t_{2}^{j,l})^{\mu}_{\nu}(\xi_{i}),\quad \tilde{F}_{\rho\lambda}^{i}=\partial_{\rho}\xi_{i\lambda}-\partial_{\lambda}\xi_{i\rho}
\end{aligned}
\end{equation}

At this stage, let us pay attentions that upon substitution of (3.3) into (3.9), these $(T^{k}_{i})^{\mu}_{\nu}$ are just the linear combinations of $(\Theta^{k})^{\mu}_{\nu}$ in (2.25) and by this reason, it is more convenient to utilize this description to deal with the issues of stability in original higher-order derived theory. Furthermore, we remark here that a simple observation shows the action functional is also equipped with the symmetries
\begin{equation}
\begin{aligned}
\delta_{\varepsilon}\xi_{i}=\varepsilon^{\mu}\partial_{\mu}(W^{k}\xi_{i})
\end{aligned}
\end{equation}
and following the same procedure employed above, one can simplify the conserved tensors as
\begin{equation}
\begin{aligned}
(T^{k}_{i})^{\mu}_{\nu}(\xi_{i})=&\sum_{l=k+1}^{p_{i}}C_{p_{i}}^{l}(-\lambda_{i})^{p_{i}-l}(\tilde{t}_{1}^{k,l})^{\mu}_{\nu}+\sum_{l=0}^{k-1}C_{p_{i}}^{l}(-\lambda_{i})^{p_{i}-l}(\tilde{t}_{2}^{k,l})^{\mu}_{\nu}\\
&+\frac{1}{2} C_{p_{i}}^{k}(-\lambda_{i})^{p_{i}-k}\delta^{\mu}_{\nu}(\square^{k-1}\partial^{\rho}\tilde{F}^{i}_{\rho\lambda}\square^{k-1}\partial_{\tau}\tilde{F}_{i}^{\tau\lambda})\\
\end{aligned}
\end{equation}

\subsection{Complex roots case}

In a similar way, for the fields $\eta_{j}$, we have the following independent higher-order symmetry transformations
\begin{equation}
\begin{aligned}
\delta_{\varepsilon}\eta_{j}=\varepsilon^{\mu}\partial_{\mu}(W^{k}\eta_{j})
\end{aligned}
\end{equation}
which are parameterized by the indices $k= 0,1,...,2q_{j}-1$. In analogy to the previous discussions, it is evident to see that the corresponding conserved quantities satisfy
\begin{equation}
\begin{aligned}
\partial_{\mu}(U^{k}_{j})^{\mu}_{\nu}=&\partial_{\nu}(W^{k}\eta_{j})(W^{2}-(\omega_{j}+\bar{\omega}_{j})W+\omega_{j}\bar{\omega}_{j})^{q_{j}}\eta_{j}\\
=&\sum_{r,s=0}^{q_{j}}C_{q_{j}}^{r,s}\partial_{\nu}(W^{k}\eta_{j})W^{2r}(-\omega_{j}-\bar{\omega}_{j})^{s}W^{s}(\omega_{j}\bar{\omega}_{j})^{q_{j}-r-s}\eta_{j}\\
\end{aligned}
\end{equation}
here
\begin{equation}
\begin{aligned}
C_{q_{j}}^{r,s}=\frac{q_{j}!}{r!s!(q_{j}-r-s)!}
\end{aligned}
\end{equation}
as a consequence, utilizing (2.25), the explicit expressions of the second-rank conserved tensors associated with complex roots take the form of
\begin{equation}
\begin{aligned}
(U^{k}_{j})^{\mu}_{\nu}(\eta_{j})=&\sum_{2r+s>k}C_{q_{j}}^{r,s}(-\omega_{j}-\bar{\omega}_{j})^{s}(\omega_{j}\bar{\omega}_{j})^{q_{j}-r-s}(t_{1}^{k,2r+s})^{\mu}_{\nu}(\eta_{j})\\
&+\sum_{2r+s<k}C_{q_{j}}^{r,s}(-\omega_{j}-\bar{\omega}_{j})^{s}(\omega_{j}\bar{\omega}_{j})^{q_{j}-r-s}(t_{2}^{k,2r+s})^{\mu}_{\nu}(\eta_{j})\\
&+\frac{1}{2}\sum_{2r+s=k}C_{q_{j}}^{r,s}(-\omega_{j}-\bar{\omega}_{j})^{s}(\omega_{j}\bar{\omega}_{j})^{q_{j}-r-s}\delta^{\mu}_{\nu}(\square^{k-1}\partial^{\rho}\bar{F}^{j}_{\rho\lambda}\square^{k-1}\partial_{\tau}\bar{F}^{\tau\lambda}_{j})
\end{aligned}
\end{equation}
here we use $\bar{F}_{\rho\lambda}^{j}=\partial_{\rho}\eta_{j\lambda}-\partial_{\lambda}\eta_{j\rho}$.

To this end, once the conserved tensors for the real and complex roots are known, by introducing two collections of parameters
\begin{equation}
\begin{aligned}
\beta_{i}^{r},\quad i=0,...,p_{i}-1,\quad r=1,...,p,\quad \gamma_{j}^{s},\quad j=0,...,q_{j},\quad s=1,...,q
\end{aligned}
\end{equation}
and summarizing all of these results together, the total conserved tensors of the lower-order action functional (3.5) can be expressed as
\begin{equation}
\begin{aligned}
\Theta^{\mu}_{\nu}=\sum_{i=1}^{p}\sum_{r=0}^{p_{i}-1}\beta_{i}^{r}(T_{i}^{r})^{\mu}_{\nu}(\xi_{i})+\sum_{j=1}^{q}\sum_{s=0}^{2q_{j}-1}\gamma_{j}^{s}(U_{j}^{s})^{\mu}_{\nu}(\eta_{j})
\end{aligned}
\end{equation}
with this result in hand, there is no difficulty in obtaining the conserved tensors of the original derived theory by inserting (3.12) and (3.16) into (3.18). Now in the consideration of the stability of the higher derivative system, the positive 00-component is relevant. In view of the independent of the fields $\xi_{i}$ and $\eta_{j}$, when $\mu=\nu=0$, the positive condition can be met only if the parameters $\beta_{i}^{r},\gamma_{j}^{s}$ fulfill
\begin{equation}
\begin{aligned}
\sum_{r=0}^{p_{i}-1}\beta_{i}^{r}(T_{i}^{r})^{0}_{0}(\xi_{i})\geq0,    \quad \quad \quad \quad \sum_{s=0}^{2q_{j}-1}\gamma_{j}^{s}(U_{j}^{s})^{0}_{0}(\eta_{j})\geq0
\end{aligned}
\end{equation}
for any solutions of $\xi_{i}$ and $\eta_{j}$. Therefore, we conclude that the derived theory is stable if and only if all the component fields are stable.

\section{Third-order derived theory}
An instructive example to illustrate the spirit in previous section is the third-order derived theory which has a three-parameter family of conserved tensors. The behavior of 00-component of the conserved tensors in such model strongly relies on the structure of the roots of the characteristic polynomial
\begin{equation}
\begin{aligned}
z^{3}+a_{2}z^{2}+a_{1}z+a_{0}=0
\end{aligned}
\end{equation}
in what follows, we consider the ansatz for different situations of the root decomposition in the third-order equation (4.1).
At first, for simplicity we restrict ourselves to the case where the polynomial possesses three different real roots $\lambda_{i}$, or in other words, the wave operator (2.6) is decomposed as follows
\begin{equation}
\begin{aligned}
M=(W-\lambda_{1})(W-\lambda_{2})(W-\lambda_{3})
\end{aligned}
\end{equation}
obviously, such situation corresponds to
\begin{equation}
\begin{aligned}
p=3,\quad \quad q=0,\quad \quad p_{i}=1
\end{aligned}
\end{equation}
in (3.1) and for every real root $\lambda_{i}$, it is convenient to define the auxiliary fields $\xi_{i}$
\begin{equation}
\begin{aligned}
\xi_{1}=(W-\lambda_{2})(W-\lambda_{3})A,\quad \xi_{2}=(W-\lambda_{1})(W-\lambda_{3})A,\quad \xi_{3}=(W-\lambda_{1})(W-\lambda_{2})A
\end{aligned}
\end{equation}
which provide the relations
\begin{equation}
\begin{aligned}
(W-\lambda_{1})\xi_{1}=0,\quad (W-\lambda_{2})\xi_{2}=0,\quad (W-\lambda_{3})\xi_{3}=0
\end{aligned}
\end{equation}
then according to (3.12), we acquire the following second-rank conserved tensors
\begin{equation}
\begin{aligned}
(T_{i})^{\mu}_{\nu}(\xi_{i})=(t^{0,1}_{1})^{\mu}_{\nu}(\xi_{i})-\frac{1}{2}\lambda_{i}\delta^{\mu}_{\nu}\xi_{i\rho}\xi_{i}^{\rho}
\end{aligned}
\end{equation}
and taking advantage of (2.26), it is not hard to write down
\begin{equation}
\begin{aligned}
(t_{1}^{0,1})^{\mu}_{\nu}(\xi_{i})=&\tilde{F}_{\nu\lambda}^{i}\tilde{F}_{i}^{\mu\lambda}-\frac{1}{4}\delta^{\mu}_{\nu}\tilde{F}^{i}_{\rho\lambda} \tilde{F}^{\rho\lambda}_{i}+\xi_{i\nu}\partial_{\tau}\tilde{F}^{\tau\mu}_{i}
\end{aligned}
\end{equation}
due to the relations (4.4), the equations of motion for $\xi_{i}$ turn out to be
\begin{equation}
\begin{aligned}
\partial_{\rho}\tilde{F}_{i}^{\rho0}-\lambda_{i}\xi_{i}^{0}=0
\end{aligned}
\end{equation}
with the aid of these equalities and in view of the metric $g_{\mu\nu}=\mathrm{diag}(1,-1,-1,-1)$, it is helpful for us to simplify the $(T_{i})^{0}_{0}$ in the form of
\begin{equation}
\begin{aligned}
(T_{i})^{0}_{0}(\xi_{i})=-\frac{1}{4}\tilde{F}_{\mu\nu}^{i}\tilde{F}^{i}_{\mu\nu}+\frac{1}{2}\lambda_{i}\xi_{i\rho}\xi_{i\rho}
\end{aligned}
\end{equation}
in this manner, the 00-component of the linear combination of $(T_{i})^{0}_{0}$ is given by
\begin{equation}
\begin{aligned}
T^{0}_{0}=&\sum_{i=1}^{3}\beta_{i}(-\frac{1}{4}\tilde{F}^{i}_{\mu\nu}\tilde{F}^{i}_{\mu\nu}+\frac{1}{2}\lambda_{i}\xi_{i\rho}\xi_{i\rho})
\end{aligned}
\end{equation}
now under the assumption
\begin{equation}
\begin{aligned}
\beta_{i}<0, \quad \quad \lambda_{i}<0 , \quad \quad  i=1,2,3
\end{aligned}
\end{equation}
the contributions of all the component fields are positive which allow us to confirm the stability of the higher derivative Abelian gauge system defined by the third-order wave operator.

Next, let us suppose the third-order equation has the simple real root $\lambda_{1}$ and real root $\lambda_{2}$ of multiplicity of 2
\begin{equation}
\begin{aligned}
M=(W-\lambda_{1})(W-\lambda_{2})^{2}
\end{aligned}
\end{equation}
which corresponds to the case of
\begin{equation}
\begin{aligned}
p=3,\quad \quad p_{1}=1, \quad \quad p_{2}=2, \quad \quad q=0
\end{aligned}
\end{equation}
in (3.1) and analogously, we define the following auxiliary fields
\begin{equation}
\begin{aligned}
\xi_{1}=(W-\lambda_{2})^{2}A,\quad\quad\quad\quad \xi_{2}=(W-\lambda_{1})A
\end{aligned}
\end{equation}
which yield the relations
\begin{equation}
\begin{aligned}
(W-\lambda_{1})\xi_{1}=0,\quad \quad \quad \quad (W-\lambda_{2})^{2}\xi_{2}=0
\end{aligned}
\end{equation}
then by virtue of formula (3.12), the second-rank conserved tensors in current situation have the form
\begin{equation}
\begin{aligned}
(T_{1})^{\mu}_{\nu}(\xi_{1})=&(t^{0,1}_{1})^{\mu}_{\nu}(\xi_{1})-\frac{1}{2}\lambda_{1}\delta^{\mu}_{\nu}\xi_{1\rho}\xi_{1}^{\rho},\\
(T^{0}_{2})^{\mu}_{\nu}(\xi_{2})=&-2\lambda_{2}(t^{0,1}_{1})^{\mu}_{\nu}(\xi_{2})+(t^{0,2}_{1})^{\mu}_{\nu}(\xi_{2})+\frac{1}{2}\lambda_{2}^{2}\delta^{\mu}_{\nu}\xi_{2\rho}\xi_{2}^{\rho},\\
(T^{1}_{2})^{\mu}_{\nu}(\xi_{2})=&(t_{1}^{1,2})^{\mu}_{\nu}(\xi_{2})+\lambda_{2}^{2}(t_{2}^{1,0})^{\mu}_{\nu}(\xi_{2})- \lambda_{2}\delta^{\mu}_{\nu}\partial^{\rho}\tilde{F}_{\rho\lambda}\partial_{\tau}\tilde{F}^{\tau\lambda}
\end{aligned}
\end{equation}
in the above expressions, the notations
\begin{equation}
\begin{aligned}
F_{\rho\lambda}=\partial_{\rho}\xi_{1\lambda}-\partial_{\lambda}\xi_{1\rho},\quad \quad \quad \tilde{F}_{\rho\lambda}=\partial_{\rho}\xi_{2\lambda}-\partial_{\lambda}\xi_{2\rho}
\end{aligned}
\end{equation}
are adopted and after a straightforward calculation of $(t^{k,l}_{i})^{\mu}_{\nu}$ in (2.26) and (2.28), we are thus led to
\begin{equation}
\begin{aligned}
(t^{0,2}_{1})^{\mu}_{\nu}(\xi_{2})=&\tilde{F}_{\nu\lambda}\square \tilde{F}^{\mu\lambda}-\frac{1}{4}\delta^{\mu}_{\nu}\tilde{F}_{\rho\lambda}\square \tilde{F}^{\rho\lambda}+\tilde{F}^{\mu\lambda}\square \tilde{F}_{\nu\lambda}-\frac{1}{4}\delta^{\mu}_{\nu}\tilde{F}_{\rho\lambda}\square \tilde{F}^{\rho\lambda}\\
&+\partial_{\lambda} \tilde{F}^{\lambda\mu}\partial^{\rho}\tilde{F}_{\rho\nu}-\frac{1}{2}\delta^{\mu}_{\nu}\partial_{\tau} \tilde{F}^{\tau\lambda}\partial^{\rho}\tilde{F}_{\rho\lambda}+\xi_{2\nu}\square\partial_{\tau}\tilde{F}^{\tau\mu},\\
(t^{1,2}_{1})^{\mu}_{\nu}(\xi_{2})=&\square \tilde{F}_{\nu\lambda}\square \tilde{F}^{\mu\lambda}-\frac{1}{4}\delta^{\mu}_{\nu}\square \tilde{F}_{\rho\lambda}\square \tilde{F}^{\rho\lambda}+\partial^{\rho}\tilde{F}_{\rho\nu}\square\partial_{\tau}\tilde{F}^{\tau\mu},\\
(t_{2}^{1,0})^{\mu}_{\nu}(\xi_{2})=&\delta^{\mu}_{\nu}\partial^{\rho}\tilde{F}_{\rho\lambda}\xi_{2}^{\lambda}-\tilde{F}_{\nu\lambda} \tilde{F}^{\mu\lambda}+\frac{1}{4}\delta^{\mu}_{\nu}\tilde{F}_{\rho\lambda}\tilde{F}^{\rho\lambda}-\xi_{2\nu}\partial_{\tau}\tilde{F}^{\tau\mu}
\end{aligned}
\end{equation}
for $\xi_{2}$, by making use of the equations of motion from (4.15)
\begin{equation}
\begin{aligned}
\square\partial_{\rho}\tilde{F}^{\rho0}-2\lambda_{2}\partial_{\rho}\tilde{F}^{\rho0}+\lambda_{2}^{2}\xi_{2}^{0}=0
\end{aligned}
\end{equation}
we are capable of writing the 00-components of $(T^{i}_{2})^{0}_{0}(\xi_{2})$ in a more compact form
\begin{equation}
\begin{aligned}
(T_{1})^{0}_{0}(\xi_{1})&=-\frac{1}{4}F_{\mu\nu}F_{\mu\nu}+\frac{1}{2}\lambda_{i}\xi_{1\rho}\xi_{1\rho},\\
(T^{0}_{2})^{0}_{0}(\xi_{2})&=-\frac{1}{2}\tilde{F}_{\mu\nu}\square \tilde{F}_{\mu\nu}+\frac{1}{2}\partial^{\mu}\tilde{F}_{\mu\rho}\partial^{\nu}\tilde{F}_{\nu\rho}+\frac{1}{2}\lambda_{2}\tilde{F}_{\mu\nu}\tilde{F}_{\mu\nu}-\frac{1}{2}\lambda_{2}^{2}\xi_{2\rho}\xi_{2\rho},\\
(T^{1}_{2})^{0}_{0}(\xi_{2})&=-\frac{1}{4}\square\tilde{ F}_{\mu\nu}\square \tilde{F}_{\mu\nu}+\lambda_{2}\partial^{\mu}\tilde{F}_{\mu\rho}\partial^{\nu}\tilde{F}_{\nu\rho}-\lambda_{2}^{2}\xi_{2\mu}\partial^{\rho}\tilde{F}_{\rho\mu}+\frac{1}{4}\lambda_{2}^{2}\tilde{F}_{\mu\nu}\tilde{F}_{\mu\nu}
\end{aligned}
\end{equation}
subsequently, by introducing a series of parameters $\beta,\beta_{0}$ and $\beta_{1}$, the total energy density of the system is given by
\begin{equation}
\begin{aligned}
T_{0}^{0}=&\beta(T_{1})^{0}_{0}+\beta_{0}(T^{0}_{2})^{0}_{0}+\beta_{1}(T^{1}_{2})^{0}_{0}\\
=&\beta(-\frac{1}{4}F_{\mu\nu}F_{\mu\nu}+\frac{1}{2}\lambda_{1}\xi_{1\rho}\xi_{1\rho})-\frac{1}{4}\beta_{1}\square \tilde{F}_{\mu\nu}\square \tilde{F}_{\mu\nu}-\frac{1}{2}\beta_{0}\tilde{F}_{\mu\nu}\square \tilde{F}_{\mu\nu}\\
&+(\frac{1}{2}\lambda_{2}\beta_{0}+\frac{1}{4}\lambda_{2}^{2}\beta_{1})\tilde{F}_{\mu\nu}\tilde{F}_{\mu\nu}+(\frac{1}{2}\beta_{0}+\beta_{1}\lambda_{2})\partial^{\mu}\tilde{F}_{\mu\rho}\partial^{\nu}\tilde{F}_{\nu\rho}\\
&-\beta_{1}\lambda_{2}^{2}\xi_{2\mu}\partial^{\rho}\tilde{F}_{\rho\mu}-\frac{1}{2}\beta_{0}\lambda_{2}^{2}\xi_{2\rho}\xi_{2\rho}
\end{aligned}
\end{equation}
it can be shown that for the field $\xi_{1}$, we simply have
\begin{equation}
\begin{aligned}
\beta<0,\quad \quad \quad \lambda_{1}<0
\end{aligned}
\end{equation}
to guarantee the stability of dynamics of $\xi_{1}$ and for the field $\xi_{2}$, to ensure the positive definite of the quadratic form, it is better to choose
\begin{equation}
\begin{aligned}
\beta_{1}<0,\quad \quad \quad \beta_{0}=-\beta_{1}\lambda_{2},\quad \quad \quad \lambda_{2}<0
\end{aligned}
\end{equation}
indeed, one can verify this assertion through the evaluation of the discriminant in the quadratic form directly. Similarly, in the case of a pair of complex conjugate roots, the linear  combination of $(U^{0}_{2})^{0}_{0}$ and $(U^{1}_{2})^{0}_{0}$ will not give us a positive conserved tensor unless the imaginary part of complex root is set to zero which turns out to be the case we discussed above.

Finally, when the third-order equation possesses real root $\lambda$ of multiplicity 3
\begin{equation}
\begin{aligned}
M=(W-\lambda)^{3}
\end{aligned}
\end{equation}
which corresponds to the case of
\begin{equation}
\begin{aligned}
p=3,\quad \quad \quad q=0,\quad \quad \quad p_{1}=3
\end{aligned}
\end{equation}
in (3.1). Then from (3.12), after a direct computation, it is not difficult to derive the explicit formulae of the conserved tensors
\begin{equation}
\begin{aligned}
(T^{0})^{\mu}_{\nu}=&3\lambda^{2}(t_{1}^{0,1})^{\mu}_{\nu}-3\lambda(t_{1}^{0,2})^{\mu}_{\nu}+(t_{1}^{0,3})^{\mu}_{\nu}-\frac{1}{2} \lambda^{3}\delta^{\mu}_{\nu}A_{\rho}A^{\rho},\\
(T^{1})^{\mu}_{\nu}=&-3\lambda(t_{1}^{1,2})^{\mu}_{\nu}+(t_{1}^{1,3})^{\mu}_{\nu}-\lambda^{3}(t_{2}^{1,0})^{\mu}_{\nu}+\frac{3}{2}\lambda^{2}\delta^{\mu}_{\nu}\partial^{\rho}F_{\rho\lambda}\partial_{\tau}F^{\tau\lambda},\\
(T^{2})^{\mu}_{\nu}=&(t_{1}^{2,3})^{\mu}_{\nu}-\lambda^{3}(t_{2}^{2,0})^{\mu}_{\nu}+3\lambda^{2}(t_{2}^{2,1})^{\mu}_{\nu}-\frac{3}{2}\lambda\delta^{\mu}_{\nu}\square\partial^{\rho}F_{\rho\lambda}\square\partial_{\tau}F^{\tau\lambda}
\end{aligned}
\end{equation}
taking into account of (2.26) and (2.28), the $(t^{k,l}_{i})^{\mu}_{\nu}$ we need in present case can be worked out in the form of
\begin{equation}
\begin{aligned}
(t_{1}^{0,3})^{\mu}_{\nu}=&F_{\nu\lambda}\square^{2} F^{\mu\lambda}-\frac{1}{4}\delta^{\mu}_{\nu}F_{\rho\lambda}\square^{2} F^{\rho\lambda}+F^{\mu\lambda}\square^{2} F_{\nu\lambda}-\frac{1}{4}\delta^{\mu}_{\nu}F_{\rho\lambda}\square^{2} F^{\rho\lambda}\\
&+\partial_{\lambda} F^{\lambda\mu}\square\partial^{\rho}F_{\rho\nu}-\frac{1}{2}\delta^{\mu}_{\nu}\partial_{\tau} F^{\tau\lambda}\square\partial^{\rho}F_{\rho\lambda}+\square F^{\mu\lambda}\square F_{\nu\lambda}\\
&-\frac{1}{4}\delta^{\mu}_{\nu}\square F_{\rho\lambda}\square F^{\rho\lambda}+\square\partial_{\lambda} F^{\lambda\mu}\partial^{\rho}F_{\rho\nu}-\frac{1}{2}\delta^{\mu}_{\nu}\square\partial_{\tau} F^{\tau\lambda}\partial^{\rho}F_{\rho\lambda}\\
&+A_{\nu}\square^{2}\partial_{\tau}F^{\tau\mu},\\
(t_{1}^{1,3})^{\mu}_{\nu}=&\square F_{\nu\lambda}\square^{2} F^{\mu\lambda}-\frac{1}{4}\delta^{\mu}_{\nu}\square F_{\rho\lambda}\square^{2} F^{\rho\lambda}+\square F^{\mu\lambda}\square^{2} F_{\nu\lambda}-\frac{1}{4}\delta^{\mu}_{\nu}\square F_{\rho\lambda}\square^{2} F^{\rho\lambda}\\
&+\square\partial_{\lambda} F^{\lambda\mu}\square\partial^{\rho}F_{\rho\nu}-\frac{1}{2}\delta^{\mu}_{\nu}\square\partial_{\tau} F^{\tau\lambda}\square\partial^{\rho}F_{\rho\lambda}+\partial^{\rho}F_{\rho\nu}\square^{2}\partial_{\tau}F^{\tau\mu},\\
(t_{1}^{2,3})^{\mu}_{\nu}=&\square^{2}F_{\nu\lambda}\square^{2} F^{\mu\lambda}-\frac{1}{4}\delta^{\mu}_{\nu}\square^{2}F_{\rho\lambda}\square^{2} F^{\rho\lambda}+\square\partial^{\rho}F_{\rho\nu}\square^{2}\partial_{\tau}F^{\tau\mu},\\
(t_{2}^{2,0})^{\mu}_{\nu}=&\delta^{\mu}_{\nu}\square\partial^{\rho}F_{\rho\lambda}A^{\lambda}-F_{\nu\lambda}\square F^{\mu\lambda}+\frac{1}{4}\delta^{\mu}_{\nu}F_{\rho\lambda}\square F^{\rho\lambda}-(F^{\mu\lambda}\square F_{\nu\lambda}\\
&-\frac{1}{4}\delta^{\mu}_{\nu}F_{\rho\lambda}\square F^{\rho\lambda}+\partial_{\lambda} F^{\lambda\mu}\partial^{\rho}F_{\rho\nu}-\frac{1}{2}\delta^{\mu}_{\nu}\partial_{\tau} F^{\tau\lambda}\partial^{\rho}F_{\rho\lambda})-A_{\nu}\square\partial_{\tau}F^{\tau\mu},\\
(t_{2}^{2,1})^{\mu}_{\nu}=&\delta^{\mu}_{\nu}\square\partial^{\rho}F_{\rho\lambda}\partial_{\tau}F^{\tau\lambda}-\square F_{\nu\lambda}\square F^{\mu\lambda}+\frac{1}{4}\delta^{\mu}_{\nu}\square F_{\rho\lambda}\square F^{\rho\lambda}-\partial^{\rho}F_{\rho\nu}\square\partial_{\tau}F^{\tau\mu}
\end{aligned}
\end{equation}
and when expanding the operator (4.24), the equation of motion for the fields $A_{\mu}$ becomes
\begin{equation}
\begin{aligned}
\square^{2}\partial_{\rho}F^{\rho0}-3\lambda\square\partial_{\rho}F^{\rho0}+3\lambda^{2}\partial_{\rho}F^{\rho0}-\lambda^{3}A^{0}=0
\end{aligned}
\end{equation}
under these constraints, the expressions of 00-components of conserved tensors are more complicate
\begin{equation}
\begin{aligned}
(T^{0})^{0}_{0}=&-\frac{3}{4}\lambda^{2}F_{\mu\nu}F_{\mu\nu}+\frac{3}{2}\lambda F_{\mu\nu}\square F_{\mu\nu}-\frac{1}{2}F_{\mu\nu}\square^{2}F_{\mu\nu}-\frac{1}{4}\square F_{\mu\nu}\square F_{\mu\nu}\\
&-\frac{3}{2}\lambda \partial^{\rho}F_{\rho\mu}\partial^{\tau}F_{\tau\mu}+\partial^{\rho}F_{\rho\mu}\square\partial^{\tau}F_{\tau\mu}+\frac{1}{2}\lambda^{3}A_{\rho}A_{\rho},\\
(T^{1})^{0}_{0}=&-\frac{1}{4}\lambda^{3}F_{\mu\nu}F_{\mu\nu}+\frac{3}{4}\lambda \square F_{\mu\nu}\square F_{\mu\nu}-\frac{1}{2}\square F_{\mu\nu}\square^{2}F_{\mu\nu}+\frac{1}{2}\square \partial^{\rho}F_{\rho\mu}\square\partial^{\tau}F_{\tau\mu}\\
&-\frac{3}{2}\lambda^{2}\partial^{\rho}F_{\rho\mu}\partial^{\tau}F_{\tau\mu}+\lambda^{3}A_{\rho}\partial^{\tau}F_{\tau\rho},\\
(T^{2})^{0}_{0}=&-\frac{1}{4}\square^{2}F_{\mu\nu}\square^{2}F_{\mu\nu}+\frac{3}{4}\lambda^{2}\square F_{\mu\nu}\square F_{\mu\nu}-\frac{1}{2}\lambda^{3}F_{\mu\nu}\square F_{\mu\nu}+\frac{1}{2}\lambda^{3}\partial^{\rho}F_{\rho\mu}\partial^{\tau}F_{\tau\mu}\\
&+\frac{3}{2}\lambda\square\partial^{\rho}F_{\rho\mu}\square\partial^{\tau}F_{\tau\mu}-3\lambda^{2}\partial^{\rho}F_{\rho\mu}\square\partial^{\tau}F_{\tau\mu}+\lambda^{3}A_{\mu}\square\partial^{\tau}F_{\tau\mu}
\end{aligned}
\end{equation}
then to fix the instability of the higher derivative system, we need parameters $\beta_{i}$ to enter the total energy density which can be illustrated as follows
\begin{equation}
\begin{aligned}
T_{0}^{0}=&\sum_{i=0}^{2}\beta_{i}(T^{i})^{0}_{0}\\
=&-\frac{1}{4}\beta_{2}\square^{2}F_{\mu\nu}\square^{2}F_{\mu\nu}+(\frac{3}{4}\lambda^{2}\beta_{2}+\frac{3}{4}\lambda\beta_{1}-\frac{1}{4}\beta_{0})\square F_{\mu\nu}\square F_{\mu\nu}-(\frac{3}{4}\lambda^{2}\beta_{0}+\frac{1}{4}\lambda^{3}\beta_{1})F_{\mu\nu}F_{\mu\nu}\\
&-\frac{1}{2}\beta_{0}F_{\mu\nu}\square^{2}F_{\mu\nu}+(\frac{3}{2}\lambda\beta_{0}-\frac{1}{2}\lambda^{3}\beta_{2})F_{\mu\nu}\square F_{\mu\nu}-\frac{1}{2}\beta_{1}\square F_{\mu\nu}\square^{2}F_{\mu\nu}\\
&+(\frac{1}{2}\beta_{1}+\frac{3}{2}\lambda\beta_{2})\square \partial^{\rho}F_{\rho\mu}\square\partial^{\tau}F_{\tau\mu}-(\frac{3}{2}\lambda\beta_{0}+\frac{3}{2}\lambda^{2}\beta_{1}-\frac{1}{2}\lambda^{3}\beta_{2})\ \partial^{\rho}F_{\rho\mu}\partial^{\tau}F_{\tau\mu}\\
&+\frac{1}{2}\lambda^{3}\beta_{0}A_{\rho}A_{\rho}+(\beta_{0}-3\lambda^{2}\beta_{2})\partial^{\rho}F_{\rho\mu}\square\partial^{\tau}F_{\tau\mu}+\lambda^{3}\beta_{2}A_{\mu}\square\partial^{\tau}F_{\tau\mu}+\lambda^{3}\beta_{2}A_{\rho}\partial^{\tau}F_{\tau\rho}
\end{aligned}
\end{equation}
now $T_{0}^{0}$ is positive and bounded only if
\begin{gather*}
\begin{pmatrix} -\frac{1}{4}\beta_{2} & -\frac{1}{4}\beta_{1} & -\frac{1}{4}\beta_{0}
\\ -\frac{1}{4}\beta_{1} & \frac{3}{4}\lambda^{2}\beta_{2}+\frac{3}{4}\lambda\beta_{1}-\frac{1}{4}\beta_{0} &\frac{3}{4}\lambda\beta_{0}-\frac{1}{4}\lambda^{3}\beta_{2}
\\ -\frac{1}{4}\beta_{0}& \frac{3}{4}\lambda\beta_{0}-\frac{1}{4}\lambda^{3}\beta_{2} &-(\frac{3}{4}\lambda^{2}\beta_{0}+\frac{1}{4}\lambda^{3}\beta_{1})
\end{pmatrix}\\
\end{gather*}
and
\begin{gather*}
\begin{pmatrix} \frac{1}{2}\beta_{1}+\frac{3}{2}\lambda\beta_{2} & \frac{1}{2}(\beta_{0}-3\lambda^{2}\beta_{2}) & \frac{1}{2}\lambda^{3}\beta_{2}
\\  \frac{1}{2}(\beta_{0}-3\lambda^{2}\beta_{2}) &-(\frac{3}{2}\lambda\beta_{0}+\frac{3}{2}\lambda^{2}\beta_{1}-\frac{1}{2}\lambda^{3}\beta_{2})& \frac{1}{2}\lambda^{3}\beta_{2}
\\  \frac{1}{2}\lambda^{3}\beta_{2}&  \frac{1}{2}\lambda^{3}\beta_{2} &\frac{1}{2}\lambda^{3}\beta_{0}
\end{pmatrix}\\
\end{gather*}
are all positive definite matrices.

\section{Conclusion and discussion}
In this paper, we investigate the stability of higher derivative Maxwell gauge field theories from the viewpoint of the $n$-parameter series of conserved quantities. These conserved quantities can be derived from the higher-order symmetries by the extension of Noether's theorem if there exists some linear operators commute with the primary wave operator. In particular, we obtain $n$ independent second-rank conserved tensors which are connected with the spacetime translation invariance of the action functional and the linear combination of these conserved tensors contains the standard canonical energy-momentum tensors. As a matter of fact, the existence of these additional conserved quantities can be seen as a consequence  of the so-called Lagrange anchor which may be traced back to the quantization of not necessarily Lagrangian dynamics~\cite{47}. In the context of general Lagrangian system or not, the Lagrange anchor maps the conserved quantities to symmetries for the field equations~\cite{48}. More importantly, it should be emphasized
that usually the Lagrange anchor is not unique in higher derivative systems and once the dynamic equations are equipped with multiple Lagrange anchors, the same symmetry can be linked to different conserved quantities. Moreover, in non-Lagrangian system, the Lagrange anchor gives us a new insight into the description of the higher derivative dynamic systems. Indeed, when the field equations admit different Lagrange anchors, the inequivalent ones will give rise to the canonically inequivalent Poisson brackets, therefore the theory turns out to be multi-Hamiltonian in the first-order formulation~\cite{49}. Especially in the derived theories, a suitable choice of parameters brings the corresponding Hamiltonian bounded from below and this classical stability can be promoted to the quantum level which implies the bounded spectrum of energy in quantum theory. Finally, the Lagrange anchor also allows us to systematically add consistent interactions into field equations of motion by proper deformation method~\cite{50} and in this sense, the conserved tensors in coupling system are regarded as the deformations of the conserved quantities in free case. It was demonstrated that if the anchor connects the symmetry with the bounded quantity, the system remains stable upon inclusion of consistent interactions. In the class of derived theories, generally speaking, the vertices of stable interactions are always non-Lagrangian but they can still admit quantization once appropriate Lagrange anchor is applied. All of these would be interesting to exploit in future.

\acknowledgments
The author would like to thank the G.W.Wan for long time encouragements and is grateful to S.M.Zhu for useful support.



\begin{thebibliography}{99}

 \bibitem{1}
M. Ostrogradsky, Mem. Acad. St. Petersbourg, VI (1850) 385.

\bibitem{2}
E.S. Fradkin and A.A. Tseytlin, \emph{Renormalizable asymptotically free quantum theory of gravity}, Nucl. Phys. B 201, 469 (1982).
\bibitem{3}
I.L. Buchbinder, S.D. Odintsov and I.L. Shapiro, \emph{Effective Action in Quantum Gravity}, IOP, Bristol,1992.
\bibitem{4}
P. Gosselin and H. Mohrbach, \emph{Renormalization of higher derivative scalar theory}, EPJ. direct 4 (2002), 1-10.
 \bibitem{5}
A. Anisimov, E. Babichev and A. Vikman,  \emph{B-inflation}, JCAP 0506: 006, (2005).
\bibitem{6}
R.P. Woodard,  \emph{Avoiding Dark Energy with $1/R$ Modifications of Gravity}, Lect. Notes. Phys, 720, 403 (2007).
\bibitem{7}
E.T. Tomboulis, \emph{Renormalization and unitarity in higher derivative and nonlocal gravity theories}, Mod.Phys.Lett. A 30 (2015), 03n04, 1540005.
\bibitem{8}
J.R. Villanueva, F. Tapia, M. Molina and M. Olivares, \emph{Null paths on a toroidal topological black hole in conformal Weyl gravity}, Eur. Phys. J. C, 78 10 (2018) 853.
\bibitem{9}
A.A. Salas, A. Molgado and E. Rojas,  \emph{Hamilton-Jacobi approach for Regge-Teitelboim cosmology}, Classical. Quant. Grav. 37 (14) 2020.

\bibitem{10}
B. Podolsky,  \emph{A generalized electrodynamics. I. Nonquantum}, Phys. Rev. (2) 62 (1942), 68–71.
\bibitem{11}
B. Podolsky and P. Schwed,\emph{Review of a Generalized Electrodynamics}, Rev. Mod. Phys. 20, 40 (1948).
\bibitem{12}
A.E.S. Green, \emph{Self-energy and interaction energy in Podolsky's generalized electrodynamics}, Phys. Rev. (2) 72 (1947), 628–631.

 \bibitem{13}
A. Pais and G.E. Uhlenbeck, \emph{On field theories with non-localized action},  Phys. Rev. 79 (1950), 145-165.
\bibitem{14}
W. Thirring, \emph{Regularization as a Consequence of Higher Order Equations},  Phys. Rev. 77 (1950), 570.



\bibitem{15}
L.V. Belvedere, R.L.P.G. Amaral and N.A. Lemos,\emph{Canonical transformations in a higher derivative field theory}, Z.Phys.C 66 (1995) 613.
\bibitem{16}
T. Nakamura and S. Hamamoto, \emph{Higher Derivatives and Canonical Formalisms}, Prog.Theor.Phys. 95 (1996) 469-484.
\bibitem{17}
F.J.de Urries and J. Julve, \emph{Ostrogradski Formalism for Higher-Derivative Scalar Field Theories}, J.Phys.A 31:6949-6964,1998.
\bibitem{18}
J. Gegelia and S. Scherer, \emph{Ostrogradsky's Hamilton formalism and quantum corrections},J.Phys.A 43 (2010) 345406.

\bibitem{19}
V.V. Nesterenko, \emph{On the instability of classical dynamics in theories with higher derivatives}, Phys.Rev.D 75 (2007) 087703.
\bibitem{20}
N.G. Stephen, \emph{On the Ostrogradski instability for higher-order derivative theories and a pseudo-mechanical energy},J. Sound. Vib 310(3):729-739，2008.
\bibitem{21}
H. Motohashi and T. Suyama,\emph{Third order equations of motion and the Ostrogradsky instability}, Phys.Rev.D 91 (2015) 8, 085009.



\bibitem{22}
H.J. Schmidt, \emph{Stability and Hamiltonian formulation of higher derivative theories}, Phys.Rev.D 49 (1994) 6354.
\bibitem{23}
A. Mostafazadeh, \emph{A  Hamiltonian formulation of the Pais–Uhlenbeck oscillator that yields a stable and unitary quantum system}, Phys.Lett.A,375(2):93-98,2010.
\bibitem{24}
M. Niedermaier,\emph{A quantum cure for the Ostrogradski instability}, Ann.Phys 327(2):329–358 ,2012.
\bibitem{25}
T. Chen, M. Fasiello, E.A. Lim and A.J. Tolley, \emph{Higher derivative theories with constraints : exorcising Ostrogradski’s ghost},  JCAP 02 (2013) 042.
\bibitem{26}
 D.S. Kaparulin and S.L. Lyakhovich, \emph{Energy and Stability of the Pais-Uhlenbeck Oscillator}, arXiv:1506.07422.
\bibitem{27}
I. Masterov, \emph{An alternative Hamiltonian formulation for the Pais-Uhlenbeck oscillator}, Nucl.Phys.B 902 (2016) 95-114.
\bibitem{28}
A. Salvio and A. Strumia, \emph{Quantum mechanics of 4-derivative theories}, Eur. Phys. J. C 76 (2016) 227.
\bibitem{29}
M.A. Camachoy, J.A. Vallejo and Y. Vorobiev, \emph{A perturbation theory approach to the stability of the Pais-Uhlenbeck oscillator}, arXiv:1703.08929.
\bibitem{30}
B. Paul, \emph{Removing Ostrogradski ghost from degenerate gravity theories}, Phys. Rev. D 96, 044035 (2017).

\bibitem{31}
M. Raidal and H. Veermae, \emph{On the Quantisation of Complex Higher Derivative Theories and Avoiding the Ostrogradsky Ghost}, Nucl. Phys. B, 916,607-626,2017.
\bibitem{32}
T.D. Lee and G.C. Wick,  \emph{Negative Metric and the Unitarity of the S Matrix},  Nucl. Phys. B 9 (1969) 209-243.
\bibitem{33}
 T.D. Lee and G.C. Wick,  \emph{Finite Theory of Quantum Electrodynamics},  Phys. Rev. D 2 (1970) 1033-1048.
 \bibitem{34}
M. Asorey, L. Rachwal and I. Shapiro, \emph{Unitary Issues in Some Higher Derivative Field Theories}, Galaxies 6 (2018) 1, 23.
\bibitem{35}
A. Mostafazadeh, \emph{Pseudo-Hermiticity versus PT symmetry 3: Equivalence of pseudoHermiticity and the presence of antilinear symmetries}, J. Math. Phys. 43, 3944 (2002).
\bibitem{36}
C.M. Bender, \emph{Introduction to PT-Symmetric Quantum Theory}, Contemp. Phys. 46, 277 (2005).
\bibitem{37}
C.M. Bender, \emph{Making sense of non-Hermitian Hamiltonians}, Rep. Prog. Phys. 70, 947 (2007).
\bibitem{38}
C.M. Bender and P.D. Mannheim, \emph{No-Ghost Theorem for the Fourth-Order Derivative Pais-Uhlenbeck Oscillator}, Phys.Rev.Lett, 100(11):110402,2007.


\bibitem{39}
D.S. Kaparulin, S.L. Lyakhovich and A.A. Sharapov,\emph{Classical and quantum stability of higher-derivative dynamics}, Eur. Phys. J. C 74(10),2014.

\bibitem{40}
D.S. Kaparulin,\emph{Conservation Laws and Stability of Field Theories of Derived Type}, Symmetry 2019, 11(5), 642.
\bibitem{41}
V.A. Abakumova, D.S. Kaparulin and S.L. Lyakhovich,\emph{Stable interactions in higher derivative field theories of derived type}, Phys. Rev. D 99, 045020,(2019).

\bibitem{42}
V.A. Abakumova, D.S. Kaparulin and S.L. Lyakhovich, \emph{Stable Interactions between extended Chern-Simons theory and charged scalar field with higher derivatives: Hamiltonian formalism}, Russ. Phys. J. 62 (2019).
\bibitem{43}
V.A. Abakumova, D.S. Kaparulin and S.L. Lyakhovich,\emph{Conservation laws and stability of higher derivative extended Chern-Simons}, arXiv:1907.02267.	
\bibitem{44}
V.A. Abakumova, D.S. Kaparulin and S.L. Lyakhovich,\emph{Stable interactions between higher derivative extended Chern-Simons and charged scalar field}, 	arXiv:1907.08075.

\bibitem{45}
D.S. Kaparulin,\emph{Lagrange Anchor for Bargmann–Wigner Equations}, arXiv:1210.2134.
\bibitem{46}
D.S. Kaparulin, S.L. Lyakhovich and A.A. Sharapov,\emph{Lagrange Anchor and Characteristic Symmetries of Free Massless Fields}, SIGMA 8 (2012), 021.


\bibitem{47}
P.O. Kazinski, S.L. Lyakhovich and A.A. Sharapov, \emph{Lagrange structure and quantization}, JHEP 0507 (2005) 076.
\bibitem{48}
D.S. Kaparulin, S.L. Lyakhovich and A.A. Sharapov, \emph{Rigid symmetries and conservation laws in non-Lagrangian field theory}, J. Math. Phys. 51 (2010) 082902.
\bibitem{49}
V.A. Abakumova, D.S. Kaparulin and S.L. Lyakhovich,\emph{Multi-Hamiltonian formulations and stability of higher-derivative extensions of 3d Chern-Simons}, Eur. Phys. J. C 78(115),2018.
\bibitem{50}
D.S. Kaparulin, S.L. Lyakhovich and A.A. Sharapov,\emph{Stable interactions via proper deformations}, J. Phys. A: Math. Theor. 49 (2016) 155204.





\end{thebibliography}
\end{document}